\documentclass[10pt,conference]{IEEEtran}
\def\BibTeX{{\rm B\kern-.05em{\sc i\kern-.025em b}\kern-.08em
    T\kern-.1667em\lower.7ex\hbox{E}\kern-.125emX}}
\usepackage{cite}
\usepackage{caption}
\usepackage{subcaption}
\usepackage{graphicx}

\usepackage{amsmath}
\usepackage{amssymb}
\usepackage{amsthm}

\usepackage{stfloats}
\usepackage{bm}
\usepackage{xcolor}

\usepackage{tikz}
\usetikzlibrary{calc,shapes,arrows,positioning,plotmarks,shadows,calc,matrix,fit,patterns}
\usepackage{pgfplots}
\usepackage{todonotes}
\usepackage{datetime}
\usepackage{cancel}
\usepackage{hhline}
\usepackage{cases}
\usepackage{nicefrac}
\usepackage{etoolbox}
\usepackage{algorithm}
\usepackage{algorithmic}
\usepackage{cleveref}
\usetikzlibrary{plotmarks}
\usetikzlibrary{arrows.meta}
\usepackage{grffile}
\usepackage{dsfont}

\newcommand{\figref}[1]{Fig.~\ref{#1}}

\newcommand{\diag}{\text{diag}}

\newcommand{\tr}{\text{tr}}

\newtheoremstyle{remarkmod}
  {\topsep}   
  {\topsep}   
  {\normalfont}  
  {0pt}       
  {\itshape} 
  {.}         
  {5pt plus 1pt minus 1pt} 
  {}          
\theoremstyle{remarkmod}

\makeatletter
\newcommand{\ALC@comblock}[1]{\ifthenelse{\equal{#1}{default}}%
{}{\textbf{#1}}}
\newenvironment{ALC@bl}{\begin{ALC@g}}{\end{ALC@g}}
\newcommand{\BLOCK}[2][default]{
	\ALC@it\ALC@comblock{#1}\ #2\begin{ALC@bl}
}
\newcommand{\ENDBLOCK}{
	\end{ALC@bl}
}

\makeatother
\usepackage{graphicx}
\usepackage{mathtools, cuted}
\usepackage{siunitx}
\IEEEoverridecommandlockouts
\begin{document}
\title{{ A Lightweight Framework for Integrated Sensing and Communications with RIS} 	
}
\author{
	\IEEEauthorblockN{Chu Li, Kevin Weinberger, Aydin Sezgin}
	\IEEEauthorblockA{Ruhr-Universit\"at Bochum, Germany\\
		Email:  \{chu.li, kevin.weinberger, aydin.sezgin\}@rub.de}
	\thanks{This work was supported in part by German Research Foundation (DFG) under the project no. 449601577, and in part by the German Research Foundation (DFG) in the course of the project SPP2433 under the project no. 541021107 (Measurement Technology on Flying Platforms)}
}
\maketitle

	\begin{abstract}
Reconfigurable Intelligent Surfaces (RIS) have been recognized as a promising technology to enhance both communication and sensing performance in integrated sensing and communication (ISAC) systems for future 6G networks. However, existing RIS optimization methods for improving ISAC performance are mainly based on semidefinite relaxation (SDR) or iterative algorithms. The former suffers from high computational complexity and limited scalability, especially when the number of RIS elements becomes large, while the latter yields suboptimal solutions whose performance depends on initialization.
In this work, we introduce a lightweight RIS phase design framework that provides a closed-form solution and explicitly accounts for the trade-off between communication and sensing, as well as proportional beam gain distribution toward multiple sensing targets. The key idea is to partition the RIS configuration into two parts: the first part is designed to maximize the communication performance, while the second introduces small perturbations to generate multiple beams for multi-target sensing. Simulation results validate the effectiveness of the proposed approach and demonstrate that it achieves performance comparable to SDR but with significantly lower computational complexity.
\end{abstract}

\section{Introduction}
RISs are metallic structures composed of a large number of low-cost passive reflective elements whose amplitudes and phases can be dynamically controlled by an attached controller. By properly configuring the RIS, the wireless propagation environment can be intelligently reconfigured, thereby enhancing both communication and sensing performance \cite{8910627,8796365,10042425}. Unlike conventional multiple-input multiple-output (MIMO) systems, which require a large number of active antennas and radio frequency (RF) chains, RISs can achieve comparable performance with significantly lower cost and power consumption.

In parallel, integrated sensing and communication (ISAC) has emerged as one of the key technologies for future 6G networks. In ISAC systems, communication and sensing share the same spectrum, hardware, and signaling resources, thereby improving spectrum utilization and hardware efficiency \cite{9540344,9354629}. Within this framework, RISs can be configured to assist both data transmission and environmental sensing without incurring additional hardware cost.
State-of-the-art studies have demonstrated the strong potential of RIS to enhance ISAC system performance. In \cite{10364760}, the authors jointly optimize the transmit beamformer and RIS configuration to maximize radar mutual information while satisfying the communication rate requirement. Due to the non-convexity of the problem, an iterative algorithm based on a one-dimensional search is employed for RIS optimization, which provides an exhaustive but computationally demanding solution. In \cite{9771801}, RIS is utilized to generate multiple beams for sensing multiple targets while ensuring that the SNR requirements at the communication receiver are met. Specifically, the worst-case beampattern gain is maximized using a semidefinite relaxation (SDR) approach. However, the SDR-based method suffers from high computational complexity, limiting its practical applicability. In \cite{9827863}, to explore the trade-off between communication and sensing, the authors maximize a weighted sum of the sensing SNR and communication SNR by jointly optimizing the transmit beamformer and RIS configuration, where a gradient descent algorithm is employed for RIS optimization. Nevertheless, iterative algorithms such as gradient descent provide only suboptimal solutions whose performance depends on the initialization point.

To address these challenges, we propose a low-complexity RIS configuration framework. The goal is to maximize communication performance while generating multiple sensing beams with proportional power allocation among targets. In addition, we explicitly investigate the trade-off between communication and sensing. To this end, the RIS configuration is divided into two components, and closed-form solutions are derived for both. Simulation results show that the proposed approach achieves performance comparable to SDR but with significantly lower computational complexity. Moreover, the proposed method can precisely control proportional power distribution among multiple targets while maintaining communication performance.



\section{System Model}
\label{sec:chmod}
     \begin{figure}[ht]
     	\centering	\resizebox{0.7\columnwidth}{!}{
     		\begin{tikzpicture}[
     			squarednode/.style={rectangle, draw=orange!60, fill=orange!40, very thick, minimum size=3mm},
     			node distance=1.5cm and 2cm,
     			box/.style={
     				draw,
     				rectangle,
     				thick,
     				inner sep=4pt,
     			}
     			]

     			\filldraw[color=green, fill=green!5, very thick]  (0,0) rectangle (3.6,2.6);
     			\node (n1) at (2,-0.25) {RIS};


     				
     				
\draw[thick] (-2.2-1,0.2-2) rectangle (-1.6-1,3.2-2);
\node[rotate=90] (tx) at (-2.9,1.7-2) {BS};
\draw[->,thick] ($(tx.east)+(1,-0.3)$) -- ($(tx.east)+(3,1.3)$);

\foreach \y in {0.6,1.3,2.0,2.7} {
	\draw[thick] (-1.6-1,\y-2) -- (-1.35-1,\y-2);
	\draw[fill=black] (-1.35-1,\y-2) -- ++(0.32,0.16) -- ++(0,-0.32) -- cycle;
}
 {};
%
%
%
%
%

     		\node[box] (rx2) at (7,1) {$\text{UE}$};
     		\draw[fill=black] ($(rx2.north west)+(0,0.2)$) -- ++(-0.15,0.3) -- ++(0.3,0) -- cycle;	    
     		\draw[thick] (rx2.north west) -- ($(rx2.north west)+(0,0.2)$);	
     			
     		\draw[->, thick]  (3.6,1.3)--($(rx2.north west)+(0,0.6)$)  node[midway, above] {} ;
     		
     		\node at (6,-1) {$\cdots$};

\node[circle, fill=red!70, inner sep=5pt] (t1) at (6,0) {};
\node[below=2pt of t1] {\small Target 1};
\draw[->, thick]  (3.6,1.3) -- (t1.west);
     		
     		\node[circle, fill=red!70, inner sep=5pt] (t2) at (5.25,-1.5) {};
     		\node[below=2pt of t2] {\small Target K};
     		\draw[->, thick]  (3.6,1.3) -- (t2.west);
     			

     			\foreach \x in {0,0.5,1,1.5,2,2.5,3} {
     				\foreach \y in {0.3,0.8,1.3,1.8,2.3} {
     					\node[squarednode] at (\x+0.3,\y) {};
     				}
     			}    			
     	\end{tikzpicture}}
     	\caption{Communication-assisted sensing system with RIS}
     	\label{fig:RIS_sys1}
     \end{figure}
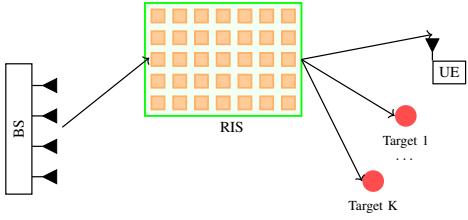

As illustrated in \figref{fig:RIS_sys1}, we consider a system where a base station (BS) equipped with $M$ antennas communicates with a single-antenna user while simultaneously sensing $K$ point targets. The direct BS-user and BS-target links are assumed to be blocked. To enable both communication and sensing, a RIS with $N$ reflecting elements is deployed to establish virtual propagation paths between the BS, the user, and the targets. In particular, we consider a communication-aided sensing framework, where the communication signals and beamforming structures are reused to sense the targets. As a result, the communication system inherently supports the sensing function without requiring additional waveforms or hardware resources, thereby achieving an efficient integration of communication and sensing \cite{10845869}.

We assume the transmitter is centered at the origin, and the RIS is deployed in close proximity to the BS so that a single line-of-sight (LoS) path exists between them. This setup is practical, as a reliable feedback link between the BS and the RIS can be easily established, allowing the RIS to be effectively controlled by the BS. We use $\mathbf{G} \in  \mathbb{C}^{M \times N}$  to denote the BS-RIS channel, which can be written as 
\begin{align}
	\label{eq:ch_G}
\mathbf{G} = \sqrt{\beta_{\mathrm{G}}}  \mathbf{g}_1  \mathbf{g}_2^H, 		
\end{align} 	
where $\beta_{\mathrm{G}}$ is the path loss factor of BS-RIS channel. Also, $ \mathbf{g}_1 \in \mathbb{C}^{M \times 1}$ represents the steering vector at the transmitter, whose $m$-th element is given by 
\begin{align}
	\mathbf{g}_{1} (m) = e^{j\frac{2\pi f_0}{c} \left( m-\frac{M+1}{2}\right) d_{\mathrm{BS}}\cos(\theta_{\mathrm{tx}})},
\end{align} 
and  $ \mathbf{g}_2 \in \mathbb{C}^{N \times 1}$ is the steering vector at the RIS, with its $n$-th element given by 
\begin{align}
	\mathbf{g}_{2} (n) \! =\! e^{-j\frac{2\pi f_0}{c}\left( \left( n_\mathrm{V} \!-\! \frac{N_\mathrm{V}+1}{2}\right) d_\mathrm{V} \sin(\theta_{\mathrm{tx}})+\left( n_\mathrm{H} \!- \!\frac{N_\mathrm{H}+1}{2}\right)d_\mathrm{H} \cos(\theta_{\mathrm{tx}})\right) },
\end{align} 
where $f_0$ denotes the carrier frequency, $c$ is the speed of light, and  $n_\mathrm{H}= \mod \left(n, N_{\mathrm{H}}\right)$ and  $n_\mathrm{V}=\left\lfloor n / N_{\mathrm{H}}\right\rfloor$ are the horizontal and vertical indices of the $n$-th element, respectively. The parameters $d_{\mathrm{BS}}$, $d_\mathrm{H}$ and  $d_\mathrm{V}$ represent the antenna spacing at the BS and the horizontal and vertical spacing of each RIS element. Furthermore, we consider the setup $d_{\mathrm{BS}} = d_\mathrm{H}= d_\mathrm{V} = \frac{c}{2 f_0}$.  Moreover, $\theta_{\mathrm{tx}}$ represents the angle of departure (AoD) at the BS. 

Unlike the BS-RIS channel, which is dominated by a deterministic LoS component, the RIS-user link often contain both LoS and non-LoS (NLoS) components. Therefore, the channel between the RIS and communication user is modeled as a Rician fading channel, given by
\begin{align}
		\label{eq:ch_RIS-UE}
	\mathbf{h}_{\text{UE}}=  \sqrt{\frac{\kappa}{1+\kappa}}  \mathbf{h}_{\text{UE}}^{\text{LOS}} +  \sqrt{\frac{1}{1+\kappa}}  \mathbf{h}_{\text{UE}}^{\text{NLOS}} , 
\end{align}
where $\kappa$ is the Rician factor. Here, $\mathbf{h}_{\text{UE}}^{\text{LOS}} \in  \mathbb{C}^{N \times 1}$ denotes the LoS component, whose $n$-th element is given by 
\begin{align}
	\mathbf{h}_{\text{UE}}^{\text{LOS}}(n)  = \sqrt{\beta_{\mathrm{UE}}} e^{-j \frac{2\pi f_0}{c} \tau_{n}^{\mathrm{RIS-UE}}},
\end{align}
where $\beta_{\mathrm{UE}}$ is the path loss factor of RIS-user channel. $\tau_{n}^{\mathrm{RIS-UE}}$ is the relative delay given by
\begin{align}
	\label{eq:delay_IRS-UE}
	\tau_{n}^{\mathrm{RIS-UE}} &=  \left( n_\mathrm{V}- \frac{N_\mathrm{V}+1}{2}\right) d_\mathrm{V} \sin(\theta_{\mathrm{UE}})
	\nonumber \\ &
	-\left( n_\mathrm{H}- \frac{N_\mathrm{H}+1}{2}\right)d_\mathrm{H} \cos(\theta_{\mathrm{UE}}),
\end{align}	
where $\theta_{\mathrm{UE}}$ is the angle of arrival (AoA) at the communication receiver.
The last term $\mathbf{h}_{\text{UE}}^{\text{NLOS}} \in \mathbb{C}^{N \times 1}$ in \eqref{eq:ch_RIS-UE} corresponds to the NLoS component, which follows a complex Gaussian distribution with zero mean and covariance $\beta_{\mathrm{UE}} \mathbf{I} $, where $\mathbf{I}$ is a $N \times N$ identity matrix.

As a communication-aided sensing framework is considered in this work, the transmitted signal $\mathbf{x} \in \mathbb{C}^{M\times1}$ is designed as
\begin{align}
	\mathbf{x}  = \mathbf{w}s, 
\end{align}
where $\mathbf{w} \in \mathbb{C}^{M\times1}$ denotes the beamforming vector. Here, $s$ is the transmission symbol for communication, which is a random variable with zero mean and unit variance. We use $P$ to denote the transmit power. Thus, we have $\mathbb{E} [\left\|	\mathbf{x}  \right\|^2 ] = \left\|\mathbf{w} \right\|^2 \leq P $. Accordingly, the received signal at the communication user is given by
\begin{align}
	\label{eq:received_signal}
	y =   \left( \mathbf{G} \diag(\mathbf{v}) \mathbf{h}_{\text{UE}}  \right) ^H \mathbf{w}s +n,
\end{align}
where $\mathbf{v} = [e^{j \phi_1},  \ldots, e^{j \phi_n}, \ldots, e^{j \phi_N}]^T$ is the phase shift vector of RIS, and $n$ is the receiver noise following a complex Gaussian distribution with zero mean and variance $\sigma^2$. 

Let $\mathbf{a}_k \in \mathbb{C}^{N\times1}$ denote the RIS steering vector corresponding to the $k$-target. Its $n$-th elements is given by
\begin{align}
\mathbf{a}_k (n) \!=\!  e^{-j \frac{2\pi f_0}{c} \left(  \left( n_\mathrm{V}\!-\! \frac{N_\mathrm{V}+1}{2}\right) d_\mathrm{V} \sin(\theta_{k})
\!-\!\left( n_\mathrm{H}\!-\! \frac{N_\mathrm{H}+1}{2}\right)d_\mathrm{H} \cos(\theta_{k})\right) },
\end{align}
where $\theta_{k}$ is the AoA of the $k$-target. Since there is no direct link between the BS and the targets, the RIS is employed to form multiple beams toward the target directions for sensing purposes. In this setting, the sensing performance can be characterized by the beampattern gain of the RIS towards the $k$-target, which is written as 
\begin{align}
	\label{eq:sensing_gain}
\mathcal{P}_k = \mathbf{a}_k^H \diag(\mathbf{v^H}) \mathbf{G}^H \mathbf{w} \mathbf{w}^H \mathbf{G} \diag(\mathbf{v}) \mathbf{a}_k.
\end{align}

\section{Beamformer and RIS Configuration}
In this section, we jointly design the transmit beamformer and the RIS configuration to enhance the received SNR at the communication receiver while also generating multiple beams toward the sensing targets. We first optimize the transmit beamformer, followed by the design of two algorithms for RIS optimization. Specifically, we derive a closed-form solution for the RIS phase-shift design and present a conventional SDR-based approach as a performance benchmark.

\subsection{Beamformer Design}
Let $\gamma$ denote the received SNR at the communication receiver. According to \eqref{eq:ch_G} and \eqref{eq:received_signal}, it is calculated as
\begin{align}
	\label{eq:snr_u}	
	\gamma &= \frac{\left|\left( \mathbf{G} \diag(\mathbf{v}) \mathbf{h}_{\text{UE}}  \right) ^H \mathbf{w}\right|^2 }{\sigma^2} \nonumber \\
	& = \frac{\beta_{\mathrm{G}}}{\sigma^2}   \left|\mathbf{h}_{\text{UE}}^H \diag(\mathbf{v}^H) \mathbf{g}_2 \mathbf{g}_1^H  \mathbf{w}\right|^2 \nonumber \\
	& =
	\frac{\beta_{\mathrm{G}}}{\sigma^2}   \left|\mathbf{h}_{\text{UE}}^H \diag(\mathbf{v}^H) \mathbf{g}_2\right|^2 \left| \mathbf{g}_1^H  \mathbf{w} \right|^2.
\end{align}
Similarly, substituting \eqref{eq:ch_G} into \eqref{eq:sensing_gain}, the beampattern gain toward the $k$-th target becomes
\begin{align}
	\label{eq:sensing_gain2}
	\mathcal{P}_k & = \beta_{\mathrm{G}}   \mathbf{w}\mathbf{a}_k^H \diag(\mathbf{v^H})\mathbf{g}_2 \mathbf{g}_1^H  \mathbf{w} \mathbf{w}^H  \mathbf{g}_1 \mathbf{g}_2^H \diag(\mathbf{v}) \mathbf{a}_k \nonumber \\
	& = \beta_{\mathrm{G}} \left| \mathbf{g}_1^H  \mathbf{w} \right|^2 \mathbf{a}_k^H \diag(\mathbf{v^H})\mathbf{g}_2 \mathbf{g}_2^H \diag(\mathbf{v}) \mathbf{a}_k.	
\end{align}
From \eqref{eq:snr_u} and \eqref{eq:sensing_gain2}, it is clear that the optimal transmit beamformer that maximizes both the received SNR at the communication user and the beampattern gain toward the sensing targets is identical, and is given by
\begin{align}
	\mathbf{w}^{*} = \sqrt{P} \frac{\mathbf{g}_1}{\left\|\mathbf{g}_1 \right\| }.
\end{align}
By plugging $\mathbf{w}^{*}$ into \eqref{eq:snr_u} and \eqref{eq:sensing_gain2}, we observe
\begin{align}
	\label{eq:snr_u2}	
	\gamma = 
	\frac{P}{\sigma^2} \beta_{\mathrm{G}} M \left|\mathbf{h}_{\text{UE}}^H \diag(\mathbf{v}^H) \mathbf{g}_2\right|^2,
\end{align}
and 
\begin{align}
	\label{eq:sensing_gain3}
    \mathcal{P}_k = P \beta_{\mathrm{G}} M \left|\mathbf{a}_k^H \diag(\mathbf{v^H})\mathbf{g}_2 \right|^2.	
\end{align}

\subsection{RIS optimization: Proposed Approach}
In this subsection, we propose a low-complexity algorithm that provides a closed-form RIS phase-shift design for SNR optimization while simultaneously generating multiple beams toward the sensing targets. The key idea is to decompose the RIS configuration into two components. As communication is considered the first priority in the system, the first component is designed to maximize the received SNR at the communication receiver, while the second introduces small phase adjustments to generate beams toward the targets.\footnote{Note that the proposed design can also be applied to the case where sensing is the first priority. In this case, the first part of the RIS can be configured to maximize the sensing performance, and small perturbations can then be introduced to enhance the communication performance.}
In this context, the RIS phase-shift vector can be expressed as
\begin{align}
	\mathbf{v} &= 	\mathbf{v}^{\star} \circ \Delta \mathbf{v} \nonumber \\ &= [e^{j \phi_1^{\star} + j\Delta \phi_1 },  \ldots, e^{j \phi_n^{\star}+j\Delta \phi_n}, \ldots, e^{j \phi_N^{\star} +j\Delta \phi_N}]^T,
\end{align}
where $\mathbf{v}^{\star} =[e^{j \phi_1^{\star} },  \ldots, e^{j \phi_n^{\star}}, \ldots, e^{j \phi_N^{\star}}]^T $ denotes the optimal phase shifts for the communication receiver and $ \Delta \mathbf{v} =[e^{j\Delta \phi_1 },  \ldots, e^{j\Delta \phi_n}, \ldots, e^{j\Delta \phi_N}]^T $
represents a small perturbation used for multi beam generation.
By substituting this expression into \eqref{eq:snr_u2}, the received SNR 
becomes
\begin{align}
	\label{eq:snr_minor_pert}	
	\gamma &= 
	\frac{P}{\sigma^2} \beta_{\mathrm{G}}    M \left| \textstyle \sum_{n=1}^{N} \mathbf{h}_{\text{UE}}^H (n) e^{-j \phi_n^{\star}-j\Delta \phi_n} \mathbf{g}_2(n)\right|^2. 
\end{align}
The RIS vector $\mathbf{v}^{\star} $ is then designed to maximize $\gamma$. Its $n$-th element is given by
 \begin{align}
 \mathbf{v}^{\star}(n) = e^{	j\phi_n^{\star}}	 =  e^{j\angle\left( \mathbf{h}_{\text{UE}}^H (n)   \mathbf{g}_2(n) \right)}. 
 \end{align}
With this choice, the beampattern gain in \eqref{eq:sensing_gain3} is calculated as
\begin{align}
	\label{eq:sensing_gain4}
	\mathcal{P}_k &= P \beta_{\mathrm{G}} M \left| \textstyle\sum_{n=1}^{N} \mathbf{a}_{k}^H (n) e^{-j \phi_n^{\star}-j\Delta \phi_n} \mathbf{g}_2(n)\right|^2 \nonumber \\
		& = P \beta_{\mathrm{G}} M \left| \textstyle\sum_{n=1}^{N}  e^{-j \left( \angle\left( \mathbf{h}_{\text{UE}} (n) \right) -  \angle\left( \mathbf{a}_{k} (n)\right)  \right)  } e^{-j\Delta \phi_n} \right|^2. 
\end{align}
Define
\begin{align}
	\eta_k = \textstyle\sum_{n=1}^{N} e^{-j\left( \angle\left(\mathbf{h}_{\text{UE}}(n)\right) - \angle \left(\mathbf{a}_k(n)\right) \right)} e^{-j\Delta \phi_n}.
\end{align}
Applying the first-order Taylor expansion $e^{x} \approx 1+x$ for small $x$, we observe
 \begin{align}
	\label{eq:snr_minor_pert2}	
		\eta_k  &= 
	\textstyle \sum_{n=1}^{N}   e^{-j \left( \angle\left( \mathbf{h}_{\text{UE}} (n) \right) -  \angle\left( \mathbf{g}_{\text{UE}_k} (n)\right)  \right)  } e^{-j\Delta \phi_n}  \nonumber \\
	& \approx 	\textstyle \sum_{n=1}^{N}   e^{-j \left( \angle\left( \mathbf{g}_{\text{UE}} (n) \right) -  \angle\left( \mathbf{g}_{\text{UE}_k} (n)\right)  \right)  } \left(1-j \Delta \phi_n\right)    \nonumber \\
	& =   \textstyle\sum_{n=1}^{N} e^{-j \left( \angle\left( \mathbf{h}_{\text{UE}} (n) \right) -  \angle\left( \mathbf{g}_{\text{UE}_k} (n)\right)  \right)  }  \nonumber \\
	&  -j \textstyle \sum_{n=1}^{N} e^{-j \left( \angle\left( \mathbf{h}_{\text{UE}} (n) \right) -  \angle\left( \mathbf{g}_{\text{UE}_k} (n)\right)  \right)  } \Delta \phi_n,	
\end{align} 
From \eqref{eq:snr_minor_pert2}, it follows that $\eta_k$ is a linear function with respect to the perturbation vector $\Delta \boldsymbol{\phi} = [\Delta \phi_1, \ldots, \Delta \phi_N]^T$.

To generate multiple beams toward the targets, we intentionally design the perturbation vector $\Delta \boldsymbol{\phi}$. To this end, we introduce a target parameter $\bar{\eta}_k$, which represents the desired value of $\eta_k$, and is assumed to be predetermined. From \eqref{eq:sensing_gain4}, it follows that, if an optimal $ \Delta \boldsymbol{\phi}^{*} $ exists that maximizes  $  \mathcal{P}_k$ for every target $k$, then we have

\begin{align}
	\label{eq:eta_UB}		
\eta_k =\textstyle	\sum_{n=1}^{N}   e^{-j \left( \angle\left( \mathbf{h}_{\text{UE}} (n) \right) -  \angle\left( \mathbf{a}_{k} (n)\right)  \right)  } e^{-j\Delta \phi_n^*} \leq N.
\end{align} 
In this context, we set 
\begin{align}
	\bar{\eta}_k = \alpha  \zeta_k	N,
\end{align}
where $0\leq\alpha \leq1$ controls the overall trade-off between the communication SNR and the sensing beampattern strength, while the normalized weights $\zeta_k$ ( $\sum_k \zeta_k = 1$) determine the proportional fairness among the sensing targets. The weights  $\zeta_k$ can be assigned based on the relative importance or priority of each target. In particular, setting equal weights yields a uniform beam power distribution, whereas adaptive weight selection enables proportional fairness, where targets with weaker channels or higher sensing accuracy requirements can be  allocated larger weights to balance the overall performance.

We now aim to design the RIS perturbation so that the beam pattern gain observed at each target approaches $ P \beta_{\mathrm{G}} M \left| 	\bar{\eta}_k \right|^2$. In this context, we define the objective function to be minimized as
\begin{align}
	\label{eq:obj}
	f(\Delta \boldsymbol{\phi}) &=\textstyle	\sum_{k  }   {P} M \beta_{\mathrm{G}}   \left| \eta_k - \bar{\eta}_k \right|^2.
\end{align}  
Furthermore, we consider
\begin{align}
	\label{eq:snr_ub}
	\mathcal{P}_k^{\text{UB}} &= {P} \beta_{\mathrm{G}} M \big|\alpha \zeta_k N\big|^2 , 
\end{align}
as an upper bound on the beam pattern gain of the proposed algorithm. This bound is achieved when the objective in \eqref{eq:obj} is minimized.

Based on these definitions, the objective in \eqref{eq:obj} can be rewritten as the quadratic form
 \begin{align}
 	\label{eq:objective}
	f(\Delta \boldsymbol{\phi}) = \left\| \sqrt{PM \beta_{\mathrm{G}}} \left( \mathbf{A} \Delta \boldsymbol{\phi} - \mathbf{b} \right) \right\| ^2,
\end{align} 
where $ \mathbf{A} \in \mathbb{C} ^{K\times N}$, whose $(k,n)$-th element is given by
 \begin{align}
 	\label{eq:matrixA}
 	 \mathbf{A} (k,n)=  e^{-j \left( \angle\left( \mathbf{h}_{\text{UE}} (n) \right) -  \angle\left( \mathbf{a}_k (n) \right) -\frac{\pi}{2} \right)  }, 
 \end{align} 
and $ \mathbf{b} \in \mathbb{C} ^{K\times 1}$, its $k$-th elements is given by
 \begin{align}
	\mathbf{b} (k) &=\textstyle \sum_{n=1}^{N} e^{-j \left( \angle\left( \mathbf{h}_{\text{UE}} (n) \right) -  \angle\left( \mathbf{a}_{{k}} (n)\right)  \right)  }  +\bar{\eta}_{k} .  
\end{align}
The least-squares solution that minimizes \eqref{eq:objective} can be directly obtained in closed form. 
However, in practice, the number of sensing targets $K$ is typically much smaller than the number of RIS reflecting elements $N$, 
and the directions of the targets may be closely spaced. 
In this case, the matrix $\mathbf{A}$ becomes highly underdetermined and its columns are strongly correlated, 
causing the matrix $\mathbf{A}^H\mathbf{A}$ in the least-squares formulation to be ill-conditioned or nearly singular. 
To address this issue and ensure numerical stability, the singular value decomposition (SVD) is employed to compute the optimal solution of $\Delta\boldsymbol{\phi}$. Furthermore, $\Delta \boldsymbol{\phi} $ is a real valued vector. In this context,
we formulate the problem of interest as
 \begin{align}  
	\text{P1:}  \min_{\Delta \boldsymbol{\phi} \in \mathbb{R}^{N} } \quad  \left\| \tilde{\mathbf{A}} \Delta \boldsymbol{\phi} - \tilde{\mathbf{b}}  \right\| ^2	+ \!	 \lambda \left\| \Delta \boldsymbol{\phi} \right\|^2, 
\end{align}
where 
\begin{align}
\tilde{\mathbf{A}} = \sqrt{PM \beta_{\mathrm{G}}}	\begin{bmatrix}
		\Re\{\mathbf{A}\} \\
	\Im\{\mathbf{A}\}
	\end{bmatrix}
, \quad
\tilde{\mathbf{b}} = \sqrt{PM \beta_{\mathrm{G}}}	\begin{bmatrix}
		\Re\{\mathbf{b}\} \\
	\Im\{\mathbf{b}\}
	\end{bmatrix},
\end{align}
with $\Re \{\cdot\}$ and $\Im\{\cdot\}$ denote the real and imaginary parts, respectively, and $\lambda$ is a regularization parameter that limits the phase perturbation.
We now decompose the matrix $\tilde {\mathbf{A}}$ as
\begin{align}
	\tilde {\mathbf{A}} = \mathbf{U}\boldsymbol{\Sigma}\mathbf{V}^H,
\end{align}
where $\mathbf{U}\in\mathbb{C}^{2K\times 2K}$ and $\mathbf{V}\in\mathbb{C}^{N\times N}$ are unitary matrices.
$\boldsymbol{\Sigma}$ is a diagonal matrix whose entries denote the singular values of $\tilde{\mathbf{A}}$. 
Using this decomposition, the closed-form solution in \eqref{eq:objective} can be expressed as
\begin{align}
	\label{eq:svd_solution}
	\Delta\boldsymbol{\phi}^{\star} 
	= \mathbf{V}\!\left(\boldsymbol{\Sigma}^2 + \lambda \mathbf{I}\right)^{-1}\!\boldsymbol{\Sigma}\mathbf{U}^H \tilde {\mathbf{b}}.
\end{align}
Note that the regularization parameter $\lambda$ also plays an important role in balancing communication performance and sensing capability. A smaller $\lambda$ allows relatively larger perturbations, which may improve sensing performance but can significantly degrade communication quality. As $\alpha$ controls the trade-off between communication and sensing, $\lambda$ should be chosen adaptively with respect to $\alpha$. Specifically, a larger $\lambda$ is used for communication-dominant scenarios (small $\alpha$) to suppress phase perturbations, whereas a smaller $\lambda$ is applied for sensing-dominant scenarios (large $\alpha$) to enable stronger beam generation toward the targets. In this work, we set
	\begin{align}
		\label{eq:reg}
		\lambda = (1-\alpha^2)\sigma_{\text{max}},
		\end{align} where $\sigma_{\text{max}}$ denotes the largest singular value of $\tilde{\mathbf{A}}$.

\subsection{RIS optimization: SDR Approach}
To employ the SDR method, we first rewrite the SNR of the communication receiver as
	\begin{align}
		\label{eq:snr_u3}	
		\gamma &= 
		\frac{P}{\sigma^2} \beta_{\mathrm{G}} M \mathbf{v}^H \left( \diag\left( \mathbf{h}_{\text{UE}}^H\right)   \mathbf{g}_2\right) \left( \diag\left( \mathbf{h}_{\text{UE}}^H\right) \mathbf{g}_2 \right)^H   \mathbf{v}   \nonumber \\ & = \tr \left(  \mathbf{V} \mathbf{\Psi}_{\text{UE}} \right), 
	\end{align}
	where $ \mathbf{V} = \mathbf{v} \mathbf{v}^H$ and 
	\begin{align}\mathbf{\Psi}_{\text{UE}} = \frac{P}{\sigma^2} \beta_{\mathrm{G}} M \left( \diag\left( \mathbf{h}_{\text{UE}}^H\right)   \mathbf{g}_2\right) \left( \diag\left( \mathbf{h}_{\text{UE}}^H\right) \mathbf{g}_2 \right)^H .  
	\end{align}
Similarly, the beam pattern gain can be written as 
	\begin{align}
	\label{eq:beam_pattern_gain}	
	\mathcal{P}_k &= 
	{P}\beta_{\mathrm{G}} M \mathbf{v}^H \left( \diag\left( \mathbf{a}_{k}^H\right)   \mathbf{g}_2\right) \left( \diag\left( \mathbf{a}_{k}^H\right) \mathbf{g}_2 \right)^H   \mathbf{v}   \nonumber \\ & = \tr \left(  \mathbf{V} \mathbf{\Psi}_{k} \right), 
\end{align}
with 
	\begin{align}\mathbf{\Psi}_{k} = {P} \beta_{\mathrm{G}} M \left( \diag\left( \mathbf{a}_{k}^H\right)   \mathbf{g}_2\right) \left( \diag\left( \mathbf{a}_{k}^H\right) \mathbf{g}_2 \right)^H .  
\end{align}
We now relax the rank one constraint of $\mathbf{V}$. Hence, the  problem of interest can be formulated as
	\begin{align}
		\text{P2:}  \quad	\max_{\mathbf{V}} \quad & \tr \left(  \mathbf{V} \mathbf{\Psi}_\text{UE} \right) \nonumber \\
		\text{s.t.} \quad &\tr \left(  \mathbf{V} \mathbf{\Psi}_k \right) \geq 	\mathcal{P}_k^{\text{Desired}},  \; k = 1, \ldots, K,  \nonumber \\
		&\mathbf{V}(n,n) = 1, \; n = 1, \ldots, N,  \nonumber \\
		&\mathbf{V} \succeq 0.
	\end{align}
The objective is to maximize the communication performance while guaranteeing that the beampattern gain requirements for all sensing targets are satisfied.
This problem is a semidefinite program (SDP) and can be solved using standard convex optimization tools.  A feasible RIS phase vector $\mathbf{v} $ can be obtained via Gaussian randomization applied to the relaxed solution \cite{wu2019intelligent}. This SDR-based method serves as a performance benchmark for evaluating the efficiency of the proposed closed-form design.

\subsection{Complexity Analysis}
The proposed SVD-based algorithm solves a regularized least-squares problem with a real-valued matrix of size $(2K \times N)$. 
Thus, the dominant singular value decomposition operation has a computational complexity of $\mathcal{O}(2K N^2)$~\cite{golub2013matrix}, 
which scales linearly with the number of targets and quadratically with the number of RIS elements. 
In contrast, the SDR approach optimizes a semidefinite matrix of size $(N \times N)$ via an interior-point method, 
whose per-iteration complexity is approximately $\mathcal{O}(N^4)$, 
and the total runtime can reach $\mathcal{O}(N^6)$ depending on the chosen convergence tolerance~\cite{5447068}. 
Thus, the proposed approach achieves a substantially lower computational cost compared with the conventional SDR method, especially for large $N$. 

\begin{figure*}[t]
	\centering
	\begin{subfigure}{0.32\textwidth}
		\includegraphics[width=\linewidth]{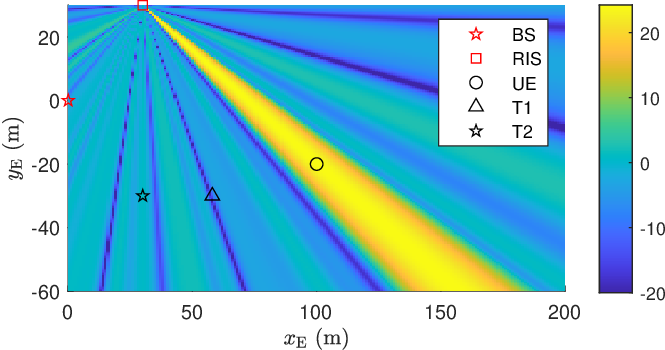}
		\caption{w/o. minor perturbations}
		\label{fig:first}
	\end{subfigure}
	\hfill
	\begin{subfigure}{0.32\textwidth}
		\includegraphics[width=\linewidth]{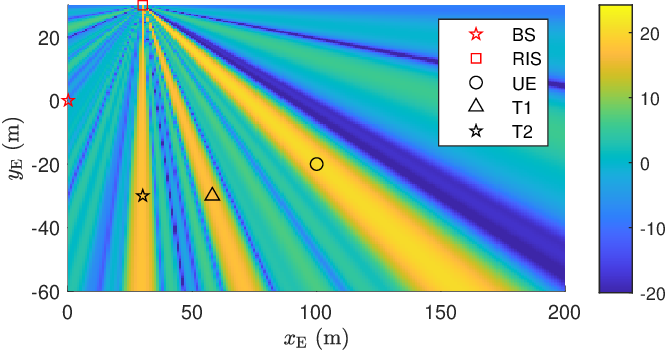}
		\caption{Proposed}
		\label{fig:second}
	\end{subfigure}
	\hfill
	\begin{subfigure}{0.32\textwidth}
		\includegraphics[width=\linewidth]{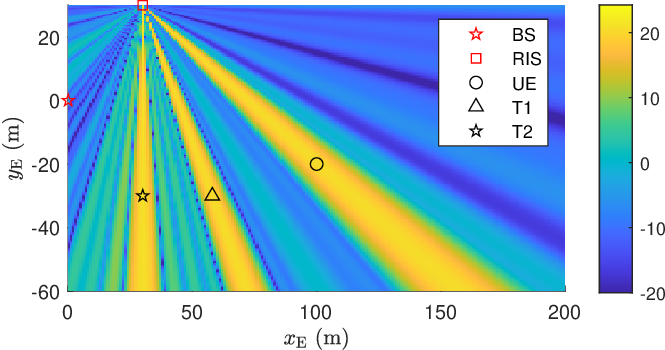}
		\caption{SDR}
		\label{fig:third}
	\end{subfigure}
	
	\caption{Heatmaps of beampattern gain with optimized RIS with $\alpha = 1$}
	\label{fig:1x3} \vspace{-1em}
\end{figure*} 
\section{Simulation Results}
\label{sec:sim_results}

In this section, we present numerical results to demonstrate the effectiveness of the proposed algorithm. Throughout the simulations, we set the transmit power to $P = 30$ dBm, and the noise power of the communication receiver to $\sigma^2 = -80$ dBm. The path loss is modeled as $\beta_{\mathrm{G}} = L_0+ 22 \log_{10} (d_1) $ [dB] and $\beta_{\mathrm{UE}} = L_0+ 25 \log_{10} (d_\mathrm{UE}) $ [dB], where $L_0 = 30$ dB is the reference path loss, $d_1$ is distance between BS and RIS, and $d_\mathrm{UE}$ is the distance between RIS and the communication receiver. The Rician factor is set to $\kappa = 1$.  Further, the center frequency is set to $f_0 = 10$ GHz, the number of transmit antennas is $M = 11$. Also, we set the number of reflective elements to $N_H = N_V = 21$. The BS is located at the origin, while the RIS is placed at $[\SI{30}{\meter}, \SI{30}{\meter}]$ and the communication receiver at $[\SI{100}{\meter}, \SI{-20}{\meter}]$.  We consider two sensing targets T1 and T2 with AoA equal 65\textdegree and 90\textdegree, unless otherwise specified.
Furthermore, to evaluate performance, the proposed approach is compared against two SDR-based benchmarks: an SDR formulation without the rank-one constraint (denoted as SDR (UB.)) and an SDR approach with 100 Gaussian randomization trials. For a fair comparison between the proposed algorithm and the SDR approach, the desired beampattern gain in the SDR case is set to be the same as that defined in \eqref{eq:snr_ub}.

In Fig.~\ref{fig:1x3}, we illustrate the spatial distribution of the beampattern gain for three RIS designs. In this example, we set $\zeta_1 = \zeta_2 = 0.5$ and $\alpha = 1$. From the figure, it can be observed that when the RIS is optimized solely for the communication receiver (Fig.~\ref{fig:first}), a narrow, high-gain beam is formed toward the user, while the targets experience very low gain (less than 0~dB). In contrast, the proposed approach (Fig.~\ref{fig:second}) can achieve stronger beampattern gains not only for the UE but also for both sensing targets (around 17~dB). The SDR-based design with Gaussian randomization (Fig.~\ref{fig:third}) exhibits a similar beampattern to the proposed approach but provides slightly higher gains at the targets (around 20~dB), while maintaining comparable gain at the UE.
\begin{figure}
	\centering
		\begin{subfigure}{0.5\textwidth}
				\includegraphics[width=0.8\linewidth]{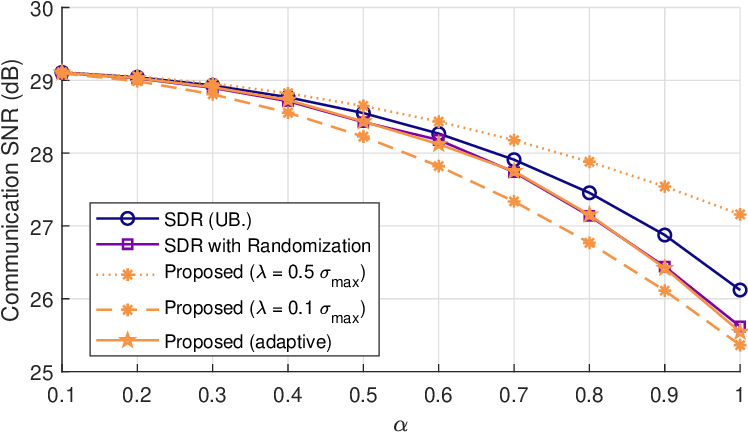}
				\caption{Communication performance with respect to $\alpha$}
				\label{fig:2a}
			\end{subfigure}
		\hfill
		\begin{subfigure}{0.5\textwidth}
		\includegraphics[width=0.8\linewidth]{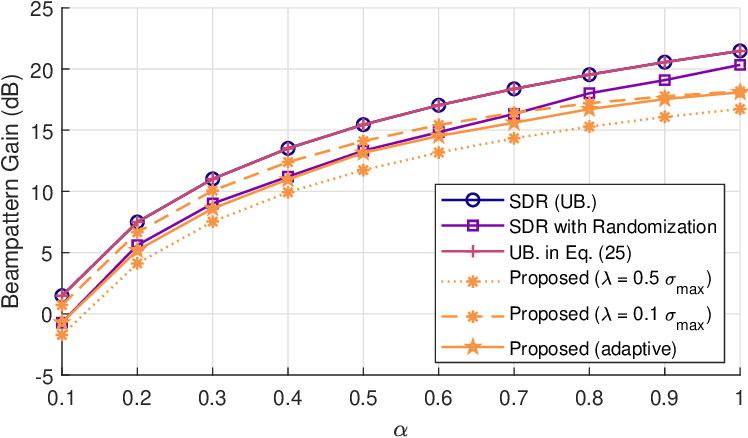}
				\caption{Sensing performance with respect to $\alpha$}
		\label{fig:2b}
			\end{subfigure}
	\caption{Comparison of communication and sensing performance with $\zeta_1=\zeta_2= 0.5$}
	\label{fig:fig2}\vspace{-2em}
\end{figure}

Fig.~\ref{fig:fig2} compares the communication and sensing performance of the proposed RIS design with different regularization parameters and the SDR-based benchmarks. As the same weight is assigned to both targets, the sensing performance for both targets is identical. Therefore, only the beampattern gain observed at T1 is plotted for clarity.
As $\alpha$ increases, the communication SNR decreases (see Fig.~\ref{fig:2a}), while the beampattern gain increases (see Fig.~\ref{fig:2b}).  For the proposed method, using a smaller $\lambda$ ($0.1\sigma_{\max}$) allows stronger perturbations, leading to improved sensing performance but degraded communication SNR. In contrast, a larger $\lambda$ ($ 0.5\sigma_{\max}$) suppresses perturbations, resulting in the highest communication SNR but the lowest sensing performance. From both figures, it can be observed that the adaptive $\lambda$ given by \eqref{eq:reg} achieves a good balance between communication and sensing. In particular, the communication SNR achieved by the proposed method with adaptive regularization closely follows the SDR performance across all values of $\alpha$. For sensing performance, the proposed method performs slightly worse at high $\alpha$ values due to the minor perturbation constraint, which limits the RIS phase adjustment and thus the achievable sensing gain.

\begin{figure}
	\centering
	\begin{subfigure}{0.5\textwidth}
		\includegraphics[width=0.8\linewidth]{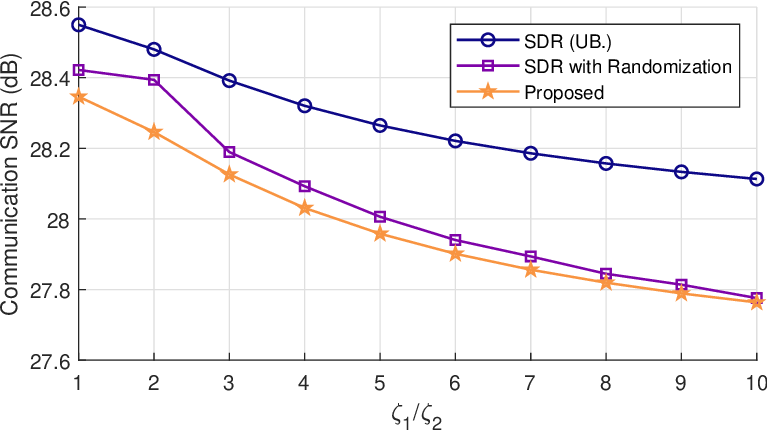}
		\caption{Communication performance with respect to $\frac{\zeta_1}{\zeta_2}$}
		\label{fig:3a}
	\end{subfigure}
	\hfill
	\begin{subfigure}{0.5\textwidth}
		\includegraphics[width=0.8\linewidth]{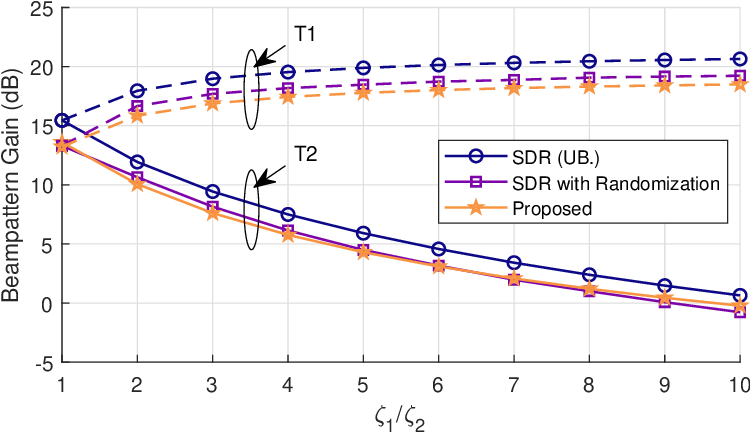}
		\caption{Sensing performance with respect to $\frac{\zeta_1}{\zeta_2}$}
		\label{fig:3b}
	\end{subfigure}
	\caption{Comparison of communication and sensing performance with $\alpha = 0.5$}
	\label{fig:fig3} \vspace{-1em}
\end{figure}

Fig.~\ref{fig:fig3} illustrates the effect of the weighting ratio $\zeta_1/\zeta_2$ on both communication and sensing performance, with $\alpha = 0.5$. The communication SNR slightly decreases as $\zeta_1/\zeta_2$ increases. This occurs because when one target is assigned a much larger weight, stronger perturbations are required to concentrate more power in that direction, thereby reducing the communication SNR. From Fig.~\ref{fig:3b}, we observe that the beampattern gain for T1 increases with $\zeta_1/\zeta_2$, while that of T2 decreases accordingly, as a larger weight is assigned to T1. Moreover, the proposed method closely follows the SDR-based benchmarks. In particular, when $\zeta_1 = \zeta_2$, both targets receive nearly equal power across all methods, whereas when $\frac{\zeta_1}{\zeta_2} = 10$, the difference reaches approximately 20~dB, as the beampattern gain scales with the square of the weights. These observations confirm that the proportional power distribution can be effectively controlled by the proposed approach.

\begin{figure}
	\centering
	\includegraphics[width=1\linewidth]{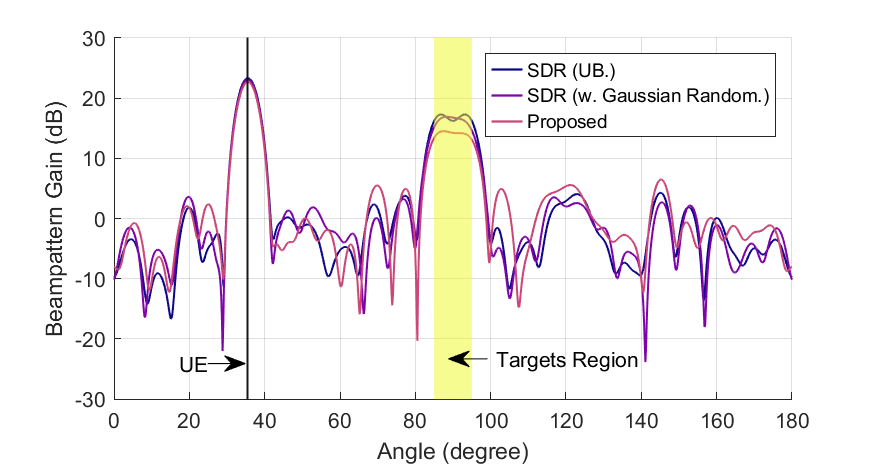}
	\caption{Comparison of Beampattern Gain with respect to AoAs with $\alpha = 0.5$}
	\label{fig:isacfig4} \vspace{-1em}
\end{figure}
Fig.~\ref{fig:isacfig4} shows the beampattern gain versus AoA for the proposed RIS design compared with the SDR-based benchmarks. Instead of assuming exact target directions, we consider targets located within the angular range $[85^\circ, 95^\circ]$. To implement both the SDR and the proposed method, the angles are uniformly sampled with an angular resolution of $1^\circ$. Both approaches form a sharp and high-gain main lobe toward the UE direction, ensuring strong communication performance. In the target region, distinct beampattern peaks can be observed, demonstrating the ability of both designs to steer energy toward the desired sensing area. Furthermore, the proposed method achieves a beampattern shape similar to that of the SDR-based approach, with only a minor reduction in peak gain. This confirms that the proposed low-complexity design can effectively achieve multi-beam formation toward both communication and sensing regions with comparable performance but significantly lower computational cost.

\section{Conclusion}
In this work, we proposed a low-complexity approach for optimizing the performance of an ISAC system. A closed-form RIS phase-shift design was derived, explicitly accounting for the trade-off between communication and sensing, as well as proportional beampattern gain allocation among multiple targets. Simulation results demonstrate that the proposed algorithm achieves performance comparable to the SDR method with Gaussian randomization, but with significantly lower computational complexity. Hence, it is well-suited for practical and real-time implementation.



\bibliographystyle{IEEEbib} 
\bibliography{refs}  
\end{document}